\documentclass{aa}

\usepackage{CJKutf8}
\usepackage{graphicx}
\usepackage[breaklinks,colorlinks,citecolor=blue,linkcolor=blue]{hyperref}
\usepackage{txfonts}
\usepackage{lscape}
\usepackage{epstopdf}
\usepackage{CJK}
\usepackage{float}
\usepackage[absolute]{textpos}
\usepackage{amsmath}
\usepackage{mathrsfs}

\usepackage{arydshln}

\usepackage{cases}

\newcommand{\xunchuancnname}{\begin{CJK*}{UTF8}{gbsn}刘训川\end{CJK*}}

\begin{document} 

   \title{Fragmentation and tidal locking in young quadruple systems}

   \author{Xunchuan Liu (\xunchuancnname)\thanks{liuxunchuan001@gmail.com} 
          }

   \institute{Leiden Observatory, Leiden University, P.O. Box 9513, 2300RA
Leiden, The Netherlands
             }

   \date{Xxx xx, 2026}
 
\abstract{
Young quadruple systems provide a particularly simple setting in which
the connection between successive levels of fragmentation can be studied.
Motivated by the recurring symmetric configurations observed in a small
number of young systems, we propose a theoretical framework in which such
configurations arise naturally during rapid collapse. Rotational
fragmentation followed by secondary fragmentation can produce two
comparable-mass components of a wide pair, each of which further
fragments into an unequal-mass close pair. The combination of tidal forces
and accretion-driven shear can then establish a preferred phase relation,
with the lower-mass component located on the inner, preceding side of the
higher-mass component in each close pair. During subsequent capture, this
phase relation can shift, placing the lower-mass component on the inner,
trailing side. Partitioning of asymmetric  accreting
flow can also tilt the spin axes of the fragments, providing a possible
origin for spin misalignments without requiring an initially incoherent
large-scale flow. The proposed mechanism therefore provides a possible
physical origin for the characteristic phase relations in young
quadruple configurations and for stochastic spin orientations within
star clusters.
}


   \maketitle

\section{Introduction}

Star formation is a multiscale process in which the structure and
dynamics of molecular clouds are progressively transformed through
gravitational collapse, fragmentation, and accretion
\citep[e.g.,][]{2007ARA&A..45..565M,2007prpl.conf..133G,2011A&A...528A..72H,2024FrASS..1103075K,2026A&A...711A.220L}.
Different spatial scales can therefore retain information about
different stages of material assembly and angular-momentum transfer
\citep[e.g.,][]{2013A&A...555A.112P,2023MNRAS.522.3719L,2025A&A...701A.141L}.
Hierarchical fragmentation provides a natural framework for connecting
these scales, as each successive fragmentation takes place within
a structure shaped by the preceding level
\citep[e.g.,][]{2013ARA&A..51..269D,2024A&A...689A.133T}.
Young quadruple systems, particularly $2+2$ configurations, represent
the simplest cases in which two distinct levels of hierarchical
structure are explicitly manifested \citep[e.g.,][]{2008MNRAS.389..925T,2021Univ....7..352T}, making them a particularly clean
laboratory for studying the early connection between successive levels.

Recent observations of young quadruple systems have revealed a
remarkably similar configuration in two systems, SSV~63 and Haro~5-2.
In SSV~63, \citet{2023AJ....165..209R} identified the multiple
protostellar system in the HH~24 complex, which was subsequently
studied in greater detail through molecular gas kinematics and
protostellar jets by \citet{2026AJ....172..108L}. A similar
configuration was reported in Haro~5-2 by
\citet{2024AJ....168..143R}. In both systems, the two ends of the
larger-scale configuration contain components of comparable mass,
while each end is further resolved into a close pair with unequal
masses. Intriguingly, the lower-mass component of each inner pair lies
on the side facing the other pair, with an angle $\theta$ of several
tens of degrees (see the sketch map in Fig.~\ref{fig:sketch}). In
SSV~63, the velocity gradient between the two ends, together with the
jet directions, suggests that the lower-mass component is on the
preceding side \citep{2026AJ....172..108L}. The recurrence of this
configuration may therefore point to an underlying dynamical process
rather than a chance arrangement, motivating the possibility that the
two hierarchical levels become dynamically coupled and maintain a
preferred phase relation during formation.

In this work, we propose a theoretical framework for the origin of such
quadruple configurations in young stellar objects, involving hierarchical
fragmentation under rotation, tidal phase locking of the fragments, and
misaligned jets from young stars embedded within the fragments. The paper
is organized as follows. In Sect.~\ref{sec_fragment}, we describe the
hierarchical rotational fragmentation picture and derive the characteristic
scales and masses of the resulting fragments. In Sect.~\ref{sec_tidall}, we
investigate how tidal forces and accretion-driven shear can establish a
phase relation between the fragments during the fragmentation and early
capture stages, when the lower-mass fragment at each end approaches its
more massive companion. Sect.~\ref{sec_discussion} discusses the
expected occurrence fraction of such configurations and the implications
of the model for misalignments between stellar spins and jets.
Sect.~\ref{sec_summary} provides a brief summary.

\section{Hierarchical fragmentation} \label{sec_fragment}
\begin{figure}[!t]
    \centering
    \includegraphics[width=0.99\linewidth]{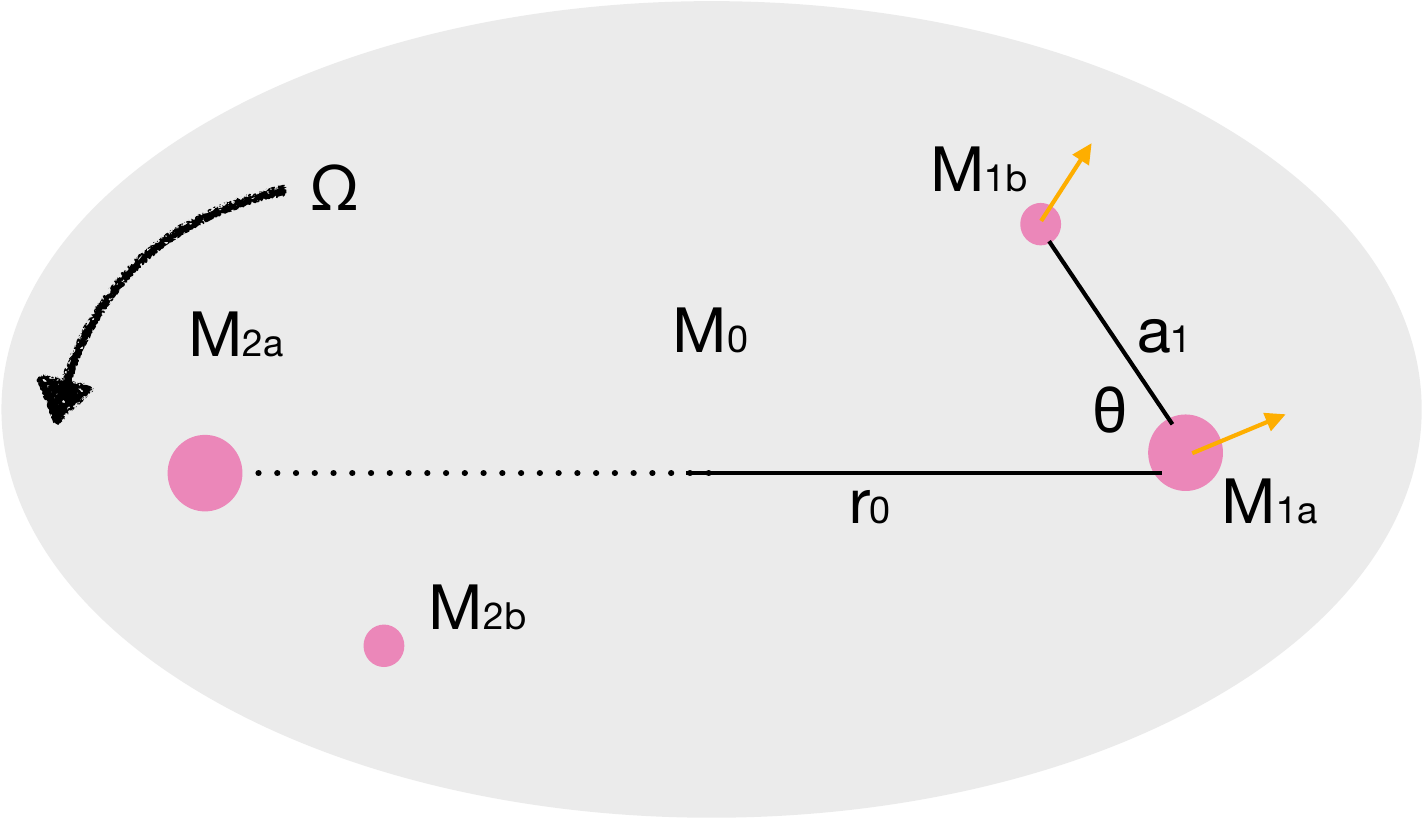}
    \caption{Schematic illustration of a phase-locked quadruple system of
young stellar objects, motivated by observations of young quadruple
systems \citep{2023AJ....165..209R,2024AJ....168..143R,2026AJ....172..108L}.
The orange arrows denote the spin directions of the inner regions
(circumstellar disks) of $M_{1a}$ and $M_{1b}$.}
    \label{fig:sketch}
\end{figure}

We consider a rapidly collapsing system, in which angular-momentum
dissipation cannot significantly alter the dynamics during the collapse,
leading to a rapidly rotating system with high angular momentum. For
simplicity, the gravitational potential is initially taken to be symmetric,
produced by a system mass $M_0$ (Fig. \ref{fig:sketch}).

\subsection{Rotational fragmentation}  \label{sec_rotfrag}

For a fluid element moving in an axisymmetric gravitational potential, the Lagrangian can be written as \citep[e.g.,][]{1999ssd..book.....M}
\begin{equation}
L=\frac{1}{2}\dot r^2+\frac{1}{2}r^2\dot\phi^2-\Phi_{\rm grav}(r),
\qquad
\Phi_{\rm grav}(r)=-\frac{GM}{r}.
\label{eq:L}
\end{equation}
The axisymmetry of $L$ implies conservation of the specific angular momentum,
\begin{equation}
j=r^2\dot\phi.
\label{eq:j}
\end{equation}
Substituting Eq.~\eqref{eq:j} into Eq.~\eqref{eq:L} gives the radial Lagrangian
\begin{equation}
L_r=\frac{1}{2}\dot r^2-\Phi_{\rm eff}(r),
\qquad
\Phi_{\rm eff}(r)=-\frac{GM}{r}+\frac{j^2}{2r^2}.
\label{eq:L_eff}
\end{equation}
The centrifugal and gravitational accelerations balance at
\begin{equation}
r_0=\frac{j^2}{GM},
\label{eq:r0}
\end{equation}
where $\Phi_{\rm eff}'(r_0)=0$.

The centrifugal barrier can therefore halt the radial contraction and produce a rotationally supported configuration, which may become susceptible to non-axisymmetric gravitational instabilities. Such perturbations can be represented as
\begin{equation}
\delta\rho\propto\cos(m\phi).
\label{eq:perturbation}
\end{equation}
For a rotating, self-gravitating configuration, the $m=2$ mode is of particular interest because it corresponds to a two-fold symmetric distortion and can produce two density enhancements on opposite sides of the rotation axis. By contrast, the $m=1$ mode corresponds primarily to a lopsided displacement, while higher-order modes produce more complex azimuthal structure. In the symmetric background considered here, the two $m=2$ density enhancements are dynamically equivalent and therefore naturally lead to comparable fragment masses,
\begin{equation}
M_1\simeq M_2.
\label{eq:equal_mass}
\end{equation}
The fragments $M_1$ and $M_2$ subsequently condense under self-gravity near $r_0$, while each remains less massive than the diffuse background component, $M_0$ (see Sect. \ref{sec_insta}). 

In this work, the fiducial values adopted for numerical estimates are
\begin{equation}
M_0=2\,M_\odot,\qquad
M_1=M_2=0.5\,M_\odot,\qquad
r_0=2000\,\mathrm{au},
\label{eq_fidu}
\end{equation}
as illustrated in Fig.~\ref{fig:sketch}.

\subsection{Instability criterion} \label{sec_insta}

The stability of the rotating system is determined by
\citep[e.g.,][]{1964ApJ...139.1217T,1992pavi.book.....S}
\begin{equation}
\nu^2=\kappa^2+\sigma^2 k^2-4\pi G\bar{\rho},
\label{eq_nu2}
\end{equation}
where $k$ is the wavenumber of the perturbation, $\kappa^2$ represents
the stabilizing effect of rotation, and $\sigma^2 k^2$ represents the
stabilizing effect of internal pressure or velocity dispersion. In
contrast, $4\pi G\bar{\rho}$ is the destabilizing self-gravity term.
The system is unstable when $\nu^2<0$. The epicyclic frequency $\kappa$ characterizes the frequency of small
radial oscillations about a circular orbit. For a Keplerian potential
\citep{1980ApJ...241..425G},
\begin{equation}
\kappa=\Omega=\sqrt{\frac{GM_c}{r^3}},
\end{equation}
where $M_c$ is the central mass.

For the whole system, the mean density is approximated as
\begin{equation}
\bar{\rho}\sim \frac{3M_0}{4\pi r_0^3}.
\end{equation}
Then, at $r=r_0$ and with $M_c=M_0$, we obtain
\begin{equation}
\nu^2
\sim
\frac{GM_0}{r_0^3}
+\sigma^2 k^2
-\frac{3GM_0}{r_0^3}
=
\sigma^2 k^2-2\Omega^2.
\end{equation}

Taking $\sigma$ to be the thermal velocity dispersion,
\begin{equation}
\sigma=\sigma_{\rm therm}
=\sqrt{\frac{k_{\rm B}T_k}{\mu m_{\rm H}}}
\simeq 0.19\,\mathrm{km\,s^{-1}},
\end{equation}
where $T_k=10\,\mathrm{K}$, $\mu=2.33$ is the mean molecular weight,
and $m_{\rm H}$ is the mass of a hydrogen atom. For the fiducial
parameters (Eq.~\ref{eq_fidu}), the corresponding angular and rotational
velocities are
\begin{equation}
\Omega
\simeq4.45\times10^{-11}\,\mathrm{s}^{-1},
\qquad
V_{\rm rot}=\Omega r_0
\simeq0.94\,\mathrm{km\,s^{-1}}.
\end{equation}
The critical wavelength is therefore
\begin{equation}
\lambda_{\rm crit}
=\frac{2\pi\sigma_{\rm therm}}{\sqrt{2}\Omega}
=\frac{2\pi\sigma_{\rm therm}}{\sqrt{2}V_{\rm rot}}r_0
\sim0.9\,r_0.
\end{equation}
Perturbations with $\lambda>\lambda_{\rm crit}$ are unstable and can
grow, whereas shorter-wavelength perturbations are stabilized by
velocity dispersion. Although the perturbations considered here are
not necessarily small, and thus $\lambda_{\rm crit}$ should not be
interpreted as an exact fragmentation scale, this linear analysis
provides a useful qualitative estimate of the characteristic
wavelength and the stability of the perturbations.

\subsection{Further fragmentation} \label{sec_secondfrag}
The characteristic sizes of the two rotational fragments, $M_1$ and
$M_2$, prior to further fragmentation and collapse, are therefore
expected to be of order
\begin{equation}
L_{M_1}\sim L_{M_2}
\sim \frac{\lambda_{\rm crit}}{2}
\sim0.5\,r_0.
\end{equation}

For $M_1$ (and similarly for $M_2$), we then use its mass and
characteristic size to estimate the mean density in Eq.~\ref{eq_nu2}.
This gives a smaller critical wavelength,
\begin{equation}
\lambda_{\rm crit,\ M_1}
\sim\lambda_{\rm crit,\ M_2}
\sim0.2\,r_0.
\end{equation}
Thus, the second level of fragmentation is favored: each of the two
rotational fragments tends to fragment into two sub-fragments, denoted
as $M_{1a}$, $M_{1b}$, $M_{2a}$, and $M_{2b}$. The tidal force (Sect. \ref{sec_tidalf}) can provide the perturbation that
drives this secondary fragmentation by stretching each rotational
fragment preferentially along the radial direction. Consequently, the
four resulting sub-fragments are expected to be distributed
approximately along a common line, before the effects of shear are
taken into account (see Sect.~\ref{sec_tidallock}).

Within $M_1$, the outer sub-fragment ($M_{1a}$) is located outside
$r_0$, whereas the inner sub-fragment ($M_{1b}$) is located inside
$r_0$. The two therefore have different access to the infalling gas.
In particular, the angular momentum of the infalling material can
prevent gas with sufficiently large specific angular momentum from
penetrating to radii smaller than $r_0$, favoring accretion by the
outer sub-fragment. Consequently, $M_{1a}$ (and similarly $M_{2a}$)
is expected to preferentially accrete the infalling gas and become
more massive than $M_{1b}$ (and $M_{2b}$).
We adopt the fiducial values (Fig. \ref{fig:sketch})
\begin{equation}
M_{1a}=0.4\,M_\odot,\qquad
M_{1b}=0.1\,M_\odot,\qquad
a_1=0.4\,r_0, \label{eq_fid12}
\end{equation}
where $a_1$ is the separation between $M_{1a}$ and $M_{1b}$.
Note that since $M_{1a} \gg M_{1b}$, $M_{1a}$ is located very close to the system radius $r_0$.

\section{Tidal locking} \label{sec_tidall}
\subsection{Tidal force}\label{sec_tidalf}
Consider the fragment $M_1$ initially located on the $x$-axis at
$r=r_0$ (Fig. \ref{fig:sketch}). The differential gravitational
acceleration across the fragment is described by the tidal acceleration
tensor \citep[see, e.g.,][]{2008gady.book.....B},
\begin{equation}
    \mathbf{a}_{\rm tid}
    =
    \mathbf{T}\boldsymbol{\ell},
    \qquad
    T_{ij}
    =
    -\frac{\partial^2\Phi}{\partial x_i\partial x_j},
\label{eq:a_tid_tensor}
\end{equation}
where $\boldsymbol{\ell}$ is the displacement from the center of $M_1$.
For the point-mass potential generated by the background component
$M_0$, the tidal tensor evaluated at $r_0$ is
\begin{equation}
    \mathbf{T}
    =
    \frac{GM_0}{r_0^3}
    \begin{pmatrix}
        2&0&0\\
        0&-1&0\\
        0&0&-1
    \end{pmatrix}.
\label{eq:tidal_tensor}
\end{equation}
Thus,
\begin{equation}
    \mathbf{a}_{\rm tid}
    =
    \frac{GM_0}{r_0^3}
    \begin{pmatrix}
        2\ell_x\\
        -\ell_y\\
        -\ell_z
    \end{pmatrix},
\label{eq:a_tid}
\end{equation}
implying stretching along the radial direction and compression in the
two transverse directions.

\subsection{Tidal locking before capture} \label{sec_tidallock}

Assume that, for $r\lesssim r_0$, the background velocity field, which is not significantly affected by the fragments, is rotationally symmetric and follows
\begin{equation}
    V_{\phi}=V_{\rm rot}\left(\frac{r}{r_0}\right)^{-\gamma}.
\end{equation}
Here, $\gamma=1/2$ corresponds to Keplerian rotation, $\gamma=1$ to a single-vortex flow, and $\gamma=-1$ to rigid-body rotation.
The local vorticity is then
\begin{equation}
    \omega_z
    = \frac{1}{r}\frac{\mathrm{d}}{\mathrm{d}r}\left(rV_{\phi}\right)
    \bigg|_{r=r_0}
    = (1-\gamma)\Omega,
\end{equation}
with an effective local angular speed
\begin{equation}
    \Omega_{\rm loc} = \frac{\omega_z}{2}
    = \frac{1-\gamma}{2}\Omega,
\end{equation}
where the factor of $1/2$ follows from the standard relation between
vorticity and the local angular velocity of a fluid element
\citep[e.g.,][]{2023SSRv..219....1T}.
For $\gamma>-1$, $\Omega_{\rm loc}<\Omega$. Thus, an object coupled to the local fluid field, e.g., through viscous coupling, would appear to rotate in the opposite direction in the frame rotating with the orbital motion. Explicitly, for a Keplerian velocity field ($\gamma=1/2$),
\begin{equation}
    \Omega_{\rm loc}=\frac{1}{4}\Omega,
\end{equation}
so that an object rotating with the local fluid field would appear to rotate in the opposite direction in the orbital frame, with a relative angular speed
\begin{equation}
    \Omega_{\rm loc,rel}
    =\Omega_{\rm loc}-\Omega
    =-\frac{3}{4}\Omega.
\end{equation}

The shear force from the background velocity field tends to push
$M_{1b}$ in the direction of the orbital motion, while the tidal force
tends to pull it back toward the radial direction. $M_{1b}$ is therefore
balanced at an angular offset $\theta$ from the radial direction (Fig. \ref{fig:sketch}). For a
Keplerian velocity field, the shear velocity relative to the orbital
frame at a radial displacement $\Delta r$ is
\begin{equation}
    \Delta V
    = r_0\frac{d\Omega}{dr}\Delta r
    = -\frac{3}{2}\Omega\Delta r.
\end{equation}
For $M_{1b}$, the radial displacement from $M_{1a}$ is
$\Delta r=a_1\cos\theta$. The shear force associated with the accretion
and pressure interaction with the background flow can then be written as
\begin{equation}
    F_{\rm shear}
    = \frac{1}{\epsilon_{\rm cross}}\dot{M}_{1b,\rm acc}\Delta V
    = -\frac{3}{2}\frac{1}{\epsilon_{\rm cross}}
    \dot{M}_{1b,\rm acc}\Omega a_1\cos\theta,
\end{equation}
where $\epsilon_{\rm cross}$ is the ratio of the accretion cross section to the
effective cross section for pressure interaction with the background
flow \citep[e.g.,][]{2001ApJ...561...69T,2024ApJ...966....7S}. The resulting torque on $M_{1b}$ about $M_{1a}$ is therefore
\begin{equation}
    \tau_{\rm shear}
    = -\frac{3}{2}\frac{1}{\epsilon_{\rm cross}}
    \dot{M}_{1b,\rm acc}\Omega a_1^2\cos^2\theta.
\end{equation}
The torque due to the tidal force (Eq.~\ref{eq:a_tid}) is
\begin{equation}
    \tau_{\rm tidal}
    = 3M_{1b}\Omega^2a_1^2\cos\theta\sin\theta. \label{eq_tautid}
\end{equation}
Torque balance, $\tau_{\rm shear}+\tau_{\rm tidal}=0$, yields
\begin{equation}
    \tan\theta
    = \frac{1}{2\epsilon_{\rm cross}}
    \frac{\dot{M}_{1b,\rm acc}}{M_{1b}}
    \frac{1}{\Omega}. \label{eqtheta_shear}
\end{equation}
Here, $\Omega^{-1}$ is the characteristic orbital timescale of the
quadruple system, while
$M_{1b}/\dot{M}_{1b,\rm acc}$ is the accretion (or collapse) timescale
during the main collapse phase of $M_{1b}$ (where
$\dot{M}_{1b,\rm acc}$ denotes the accretion rate of the collapsing core,
rather than that of the embedded protostar). Thus,
\begin{equation}
    \tan\theta
    = \frac{1}{2\epsilon_{\rm cross}}
    \frac{t_{\rm orbit}}{t_{\rm acc}},
\end{equation}
where $t_{\rm orbit}\equiv\Omega^{-1}$ and
$t_{\rm acc}\equiv M_{1b}/\dot{M}_{1b,\rm acc}$.

The typical value of $t_{\rm orbit}$ is $\sim10^4$ yr.
A substantial offset angle requires the accretion timescale during the
main collapse phase to be comparable to, or shorter than, the orbital
timescale. Under the fiducial parameters adopted in
Eqs.~\ref{eq_fidu} and \ref{eq_fid12}, this condition implies a peak
accretion rate of order
$\dot{M}_{1b,\rm acc}\sim10^{-5}\,M_\odot\,{\rm yr}^{-1}$,
which is comparable to the accretion rates expected during the early Class 0 phase of low-mass protostellar evolution \citep[e.g.,][]{1996A&A...311..858B,2014ApJ...797...32P}.

The phase-locking mechanism of a fragment pair (Sect.~\ref{sec_tidallock})
is analogous in spirit to tidal locking in planetary systems
\citep[e.g.,][]{1999ssd..book.....M}. In both cases, differential forces
acting across the system generate torques that modify the relative motion
and tend to establish a preferred phase relationship. The underlying
physical mechanisms, however, are distinct. Conventional tidal locking
is driven by differential gravitational forces coupled with internal
dissipation, whereas the phase locking considered here is driven by
differential momentum deposition through accretion from the surrounding
gas.

\subsection{Tidal locking during initial capture} \label{sec_tidallock_aftercap}
The relative motion of the two
sub-fragments along the $a_1$ direction, i.e., along the line connecting
$M_{1b}$ and $M_{1a}$, is constrained by gravity and, during the early
collapse phase, may also be influenced by the pressure gradient between
the two sub-fragments. Consequently, the subsequent evolution of the
pair is not uniquely determined by the phase-locking process itself.
Depending on the balance between gravity, pressure, and the accretion
history, the two sub-fragments may either remain separated or evolve
toward the capture of $M_{1b}$ by $M_{1a}$. Determining which outcome
occurs requires following the coupled dynamical and accretion evolution
of the system, given the intrinsically chaotic nature of few-body
gravitational dynamics \citep[e.g.,][]{2024MNRAS.528..198B}.

Assuming that $M_{1b}$ is eventually captured by $M_{1a}$, we consider
the inward motion of $M_{1b}$ toward $M_{1a}$, during which the
separation $a_1$ gradually decreases as the pressure support between
$M_{1a}$ and $M_{1b}$ weakens. In the orbital frame, the radial
velocity of $M_{1b}$ during the initial capture gives rise to a
tangential Coriolis force,
\begin{equation}
    F_{\rm capture}
    = 2M_{1b}\Omega V_r
    = -2M_{1b}\Omega\dot{a}_1.
\end{equation}
The resulting torque on $M_{1b}$ within the $M_{1b}$--$M_{1a}$ system is
\begin{equation}
    \tau_{\rm capture}
    = -2 M_{1b} \Omega a_1 \dot{a}_1
    = 2 M_{1b} \Omega a_1 |\dot{a}_1|,
\end{equation}
where $\dot{a}_1<0$ during the inward motion, and hence
$\tau_{\rm capture}>0$, i.e., the torque acts in the direction of the
orbital angular momentum. Here,
\begin{equation}
    t_{\rm capture}
    \equiv -\frac{a_1}{\dot{a}_1}
\end{equation}
is the characteristic capture timescale. During the initial capture
stage, $t_{\rm capture}$ is expected to be of the order of the local
dynamical timescale around $M_{1a}$, as $M_{1b}$ gradually decouples
from the global orbit and becomes increasingly governed by the
gravitational potential of $M_{1a}$. During this stage, the magnitude
of $\tau_{\rm capture}$ may become comparable to that of
$\tau_{\rm tidal}$ (Eq.~\ref{eq_tautid}). Their competition can
therefore produce a finite phase offset, with $M_{1b}$ locked behind
$M_{1a}$, corresponding to a negative phase offset $\theta$
(Fig.~\ref{fig:sketch}),
\begin{equation}
    \sin(2\theta)
    = -\frac{4}{3}
    \frac{t_{\rm orbit}}{t_{\rm capture}}.
    \label{eqtheta_capture}
\end{equation}
As the capture proceeds, $a_1$ decreases and the capture torque may
become increasingly important. Eventually, $M_{1b}$ is fully captured
by $M_{1a}$, and its motion becomes primarily governed by the
gravitational potential of $M_{1a}$ rather than by the global orbital
motion. The phase-locked configuration is then no longer maintained,
and the phase symmetry between $M_{1b}$ and $M_{1a}$ rapidly disappears.

\section{Discussion} \label{sec_discussion}
\subsection{Occurrence fraction}\label{sec_occurrence}
We suggest that $\theta$ (Fig.~\ref{fig:sketch}) may serve as an
evolutionary indicator during the fragmentation, initial collapse, and
early capture of a minor fragment by a companion orbiting within a global
potential field (see Eqs.~\ref{eqtheta_shear} and
\ref{eqtheta_capture} in Sect.~\ref{sec_tidall}). A quadruple system
provides a particularly favorable configuration for this purpose, as
the presence of companions at two approximately symmetric positions
provides a clear geometric signature of phase locking and allows the
phase offset $\theta$ to be directly constrained. However, this
evolutionary stage is expected to be short-lived because of the strong
dynamical instability of quadruple systems. In a rapidly
collapsing system, the duration of this stage may be only several
$10^4$ yr; we adopt a fiducial value of
$t_{\rm lock}\sim5\times10^4$ yr.

The short duration of this phase also implies that such configurations
should be rare observationally. Treating the phase-locked configuration
as a specific evolutionary stage of the broader quadruple population,
we estimate its occurrence fraction using a simple duty-cycle argument,
\begin{equation}
    f_{\rm lock}
    \sim
    f_{\rm quad}
    \frac{t_{\rm lock}}{t_{\rm Class\,I}}
    \sim
    0.04
    \frac{5\times10^4}{5\times10^5}
    \sim
    4\times10^{-3}.
\end{equation}
Here, $f_{\rm quad}\simeq0.04$ is motivated by the empirical
multiplicity statistics of solar-type stars, for which quadruple
systems constitute approximately $4\%$ of the population
\citep{2014AJ....147...87T,2024A&A...690A.385D}. The adopted
$t_{\rm Class\,I}\sim5\times10^5$ yr is a conservative characteristic
timescale for the Class~I phase, consistent with observational estimates
of several $10^5$ yr \citep[e.g.,][]{2009ApJS..181..321E}. This estimate should
be regarded as an upper-level order-of-magnitude value, since only a
subset of quadruple systems are expected to undergo the specific
wide, high-angular-momentum phase considered here. Within $\sim1$ kpc, where such quadruple systems with symmetric
configurations can be relatively easily resolved, Spitzer surveys identify
a nearby Class~I population of order $10^3$
\citep{2009ApJS..184...18G}. Combined with the estimated occurrence
fraction, this suggests that only a few such systems may be observable
in the nearby Class~I population.

At present, quadruple systems with symmetric configurations are most
readily identified in nearby star-forming regions
\citep{2023AJ....165..209R,2024AJ....168..143R,2026AJ....172..108L}.
Interferometric facilities such as ALMA can potentially enlarge the
observable sample by enabling high-resolution observations of more
distant, massive star-forming regions
\citep[e.g.,][]{2024RAA....24b5009L,2026ApJ..1006...25Y}. However,
whether clustered environments preferentially produce such systems
remains unclear. Although these environments may provide higher angular
momentum, which can promote fragmentation into multiple components
\citep[e.g.,][]{2007prpl.conf..133G}, the complex gravitational
interactions and dynamical perturbations within dense clumps may disrupt
the phase alignment in such fragile, wide configurations
\citep{2024A&A...690A.385D,2026A&A...711A.220L}.

\subsection{Spin misalignment}\label{sec_misalignment}

The accretion onto the mid-plane of the whole system may be asymmetric
between the upper and lower sides. Although accretion at different
locations can balance the torque within the plane, it may still produce
a net torque on the spin of $M_1$. This is because material in the two
half-planes separated by the axis connecting $M_1$ and $M_2$ may have
different preferences for being accreted by $M_1$ and $M_2$, resulting
in a net angular momentum that can alter the spin direction of $M_1$,
and similarly that of $M_2$. We denote the resulting torque as
\begin{equation}
    \tau_{\rm asym}
    =
    \epsilon_{\rm asym}
    \dot{M}_{\rm 0,acc}r_0^2\Omega,
\end{equation}
where $\epsilon_{\rm asym}\ll1$ is a dimensionless coefficient that
accounts for the degree of asymmetry and the ratio of the perpendicular
to tangential velocities in the global accretion flow.

For $M_1$, the resulting torque is directed outward along $r_0$,
which tends to tilt the spin axis of $M_1$ away from the direction
perpendicular to the orbital plane. A natural characteristic angular
momentum associated with this torque can be obtained by assuming that
the torque is balanced by the precession of an outward-pointing angular
momentum at a rate of order $\Omega$,
\begin{equation}
    L_{\rm tilt}
    =
    \frac{\tau_{\rm asym}}{\Omega}
    =
    \epsilon_{\rm asym}
    \dot{M}_{\rm 0,acc}r_0^2.
\end{equation}
At the characteristic size $a_1$, corresponding to the size of $M_1$
before collapse, this angular momentum corresponds to an effective
angular velocity
\begin{equation}
    \omega_{\rm tilt}
    \sim
    \frac{L_{\rm tilt}}{M_1a_1^2}
    =
    \epsilon_{\rm asym}
    \frac{r_0^2}{a_1^2}
    \frac{\dot{M}_{\rm 0,acc}}{M_1}.
\end{equation}
During the initial collapse phase, the global accretion rate can be
high, with $\dot{M}_{\rm 0,acc}/M_1$ comparable to $\Omega$. Moreover,
the large fragment size ratio $r_0/a_1$ can compensate for the small
coefficient $\epsilon_{\rm asym}$. Thus, $\omega_{\rm tilt}$ can also
be comparable to $\Omega$, allowing the disks embedded in $M_1$, which
have decoupled from the orbit of $M_1$, to preserve the tilt imprinted
during the initial collapse.

Extending this picture from two-level to multi-level hierarchical
fragmentation, the combination of tidal locking and spin tilting may
therefore partly explain the observed diversity of orientations in
young stellar systems, including misalignments between filaments and
young-star jets or stellar spins, as well as between the jet or spin
directions of different stars within a cluster
\citep[e.g.,][]{2010MNRAS.402.1380J,2017ApJ...846...16S,2017NatAs...1E..64C,2018MNRAS.479..391K,2020ApJ...896...11F,2026AJ....172..108L}.
More fundamentally, the stochastic nature of these misalignments need
not originate entirely from turbulence. It can also arise from the
chaotic partitioning of accreting mass among different fragments,
providing a mechanism for generating stochastic spin directions from
a globally coherent accretion flow.

\section{Summary} \label{sec_summary}

We have proposed a theoretical framework for the formation and early
evolution of young quadruple systems with the symmetric configuration
motivated by recent observations. Rapid rotation during collapse can
drive hierarchical fragmentation, with a first-level of fragmentation
producing two comparable-mass fragments, each of which can subsequently
undergo secondary fragmentation. The resulting fragments can develop a
preferred phase relation, with the lower-mass fragment locked on the
inner, preceding side of the higher-mass fragment in each close pair,
through the competition between tidal forces and accretion-driven shear.
During the early capture stage, if the lower-mass fragment is eventually
captured by the higher-mass fragment, the phase relation can shift, with
the lower-mass fragment becoming locked on the inner, trailing side.
Global asymmetric accretion can further tilt the spin axes of the
fragments, potentially producing misaligned stellar jets. The short
lifetime of the phase-locked configuration implies a low occurrence
fraction among Class~I systems, with our fiducial estimate giving
$f_{\rm lock}\sim4\times10^{-3}$. These results suggest that the observed
recurrence of similar quadruple configurations may reflect an underlying
dynamical process linking hierarchical fragmentation, orbital phase
relations, and the spin and jet orientations of young stars.

\begin{acknowledgement}
X. Liu acknowledges the support of the Strategic Priority Research Program of the Chinese Academy of Sciences  under Grant No. XDB0800303.
\end{acknowledgement}

\bibliographystyle{aa}
\bibliography{tidalLockYSOs}

\end{document}